\documentclass[preprint,12pt]{elsarticle}

\usepackage{graphicx}
\usepackage{longtable}
\usepackage{hyperref}
\usepackage{lineno}
\usepackage{natbib}
\usepackage{geometry}
\usepackage{fleqn}
\usepackage{txfonts}
\usepackage{color}

\journal{Signal Processing}

\begin{document}

\begin{frontmatter}

\title{Filter Design for the Detection/Estimation of the
    Modulus of a Vector. \\
    Application to Polarization Data}

\author[label1]{Francisco Arg\"ueso}
\author[label2]{Jos\'e Luis Sanz}
\author[label2]{Diego Herranz} %\corref{cor1}
\ead{herranz@ifca.unican.es}

\address[label1]{Departamento de Matem\'aticas, Universidad de Oviedo,
  33007, Oviedo, Spain}
\address[label2]{Instituto de F\'\i{sica} de Cantabria (CSIC-UC),
  39005, Santander, Spain}


\begin{abstract}
We consider a set of $M$ images, whose pixel intensities at a common
point can be treated as the components of a $M$-dimensional vector.
We are interested in the estimation of the modulus of such a vector
associated to a compact source. For instance, the
detection/estimation of the polarized signal of compact sources
immersed in a noisy background is relevant in some fields like
Astrophysics. We develop two different techniques, one based on the
Maximum Likelihood Estimator (MLE) applied to the modulus
distribution, the modulus filter (ModF) and other based on
prefiltering the components before fusion, the filtered fusion (FF),
to deal with this problem. We present both methods in the general
case of $M$ images and apply them to the particular case of three
images (linear plus circular polarization). Numerical simulations
have been performed to test these filters considering polarized
compact sources immersed in stationary noise. The FF performs better
than the ModF in terms of errors in the estimated amplitude and
position of the source, especially in the low signal-to-noise case.
We also compare both methods with the direct application of
a matched filter (MF) on the polarization data. This last technique
is clearly outperformed by the new methods.
\end{abstract}


\end{frontmatter}

\section{Introduction}

The detection and estimation of the intensity of compact objects
--i.e. signals with a compact support either in time or space
domains-- embedded in a background plus instrumental noise is a
problem of interest in many different areas of science and
engineering. A classic example is the detection of point-like
extragalactic objects such as galaxies and galaxy clusters in
sub-millimetric Astronomy. Regarding this particular field of
interest, different techniques have proven useful in the literature.
Some of the proposed techniques are frequentist, such as the standard
matched filter \cite{MF_radio92}, the matched multifilter
\cite{herr02a} or the recently developed matched matrix filters
\cite{herranz08a}. Other frequentist techniques include continuous
wavelets like the standard Mexican Hat \cite{wsphere} and other
members of its family \cite{MHW206} and, more generally, filters based
on the Neyman-Pearson approach using the distribution of maxima
\cite{can05b}. All these filters have been applied to real data of the
Cosmic Microwave Background (CMB), like those obtained by the WMAP
satellite \cite{NEWPS07} and CMB simulated data \cite{challenge08} for
the experiment on board the \emph{Planck} satellite
\cite{planck_tauber05}, which has recently started
operations. Besides, Bayesian methods have also been recently
developed \cite{psnakesI}. Although we have chosen the particular case
of sub-millimetric Astronomy as a means to illustrate the problem of
compact source detection, the methods listed above are totally general
and can be used to any analogous image processing problem.

In most cases one is interested only in the intensity of the compact
sources. In other cases, however, there are other properties of the
signal that may be of interest. Such is the case, for example, of
sources that emit electromagnetic radiation that is at least partially
polarized. Polarization of light is conventionally described in terms of the Stokes parameters
$Q$, $U$ and $V$.  Let us consider a monochromatic, plane electromagnetic wave propagating in the $z$-direction. The components of the wave's electric field vector at a given point in the space can be written as
\begin{eqnarray}
E_x & = & a_x(t) \cos \left[ \omega_0 t - \theta_x(t) \right] \nonumber \\
E_y & = & a_y(t) \cos \left[ \omega_0 t - \theta_y(t) \right].
\end{eqnarray}
\noindent
If some correlation exists between the two components in the previous
equation, then the wave is \emph{polarized}.  The Stokes parameters
are defined as the time averages
\begin{eqnarray}
I & \equiv & \langle a_x^2 \rangle + \langle a_y^2 \rangle , \nonumber
\\ Q & \equiv & \langle a_x^2 \rangle - \langle a_y^2 \rangle ,
\nonumber \\ U & \equiv & \langle 2 a_x a_y \cos \left( \theta_x -
\theta_y \right) \rangle , \nonumber \\ V & \equiv & \langle 2 a_x a_y
\sin \left( \theta_x - \theta_y \right) \rangle .
\end{eqnarray}

The parameter $I$ gives the intensity of the radiation which is always
positive, while the other three parameters define the polarization
state of the wave and can have either sign. $Q$ and $U$ are the linear
polarization parameters and $V$ indicates the circular polarization of
the wave. Unpolarized radiation is described by $Q=U=V=0$.

While the total intensity of the wave is independent of the
orientation of the $x$ and $y$ axes, the values of the other Stokes
parameters are not invariant with respect to changes of the
orientation of the receivers. On the other hand, the \emph{total
  polarization}, defined as
\begin{equation} \label{eq:P}
P \equiv \sqrt{Q^2+U^2+V^2},
\end{equation}
\noindent
is invariant with respect to the relative orientation of the receivers
and the direction of the incoming light, and therefore it is a
quantity with a clear physical meaning.  For the case of purely linear
polarization, $V=0$ and the previous expression reduces to $P \equiv
(Q^2+U^2)^{1/2}$. Note that in order to get $P$ from its components
$Q$, $U$, $V$ it is necessary to perform a non-linear operation.

Although strictly speaking $Q$, $U$ and $V$ are the components of a
tensor, the invariant combination (\ref{eq:P}) gives a quantity that
can be seen as the modulus of a vector. In this paper we will
introduce a methodology that can be applied to any problem in which
a set of images contain signals whose individual intensities can be
considered as the components of a vector, but where the quantity of
interest is the modulus of such a vector. For illustrative purposes,
throughout the paper we will use as an example the case of light
polarization, but the methods we will introduce are not limited to
the example. Another possible application could be the determination
of the modulus of a complex-valued signal. For example, in
\cite{oberto} the case of the estimation of the modulus of a
complex-valued quantity whose components follow a Gaussian
distribution was addressed. The techniques presented in their paper
are related to our methods, but restricted to the two-dimensional
case. In \cite{simmons85}, four methods: MLE estimator, median
estimator, mean estimator and Wardle and Kronberg's estimator \cite{wardle74}
are applied to the estimation of polarization (two-dimensional
case). However, these methods are applied in the cited paper to a
single data and cannot be, except for the MLE, generalized to the
detection of a signal with a given profile in a pixelized image, as
considered in our paper. This is due to the fact that these methods
lead to a system of possibly incompatible equations.

Going back to the case of point-like extragalactic objects in
sub-millimetric Astronomy, the polarization of the sources plays and
important role in the cosmological tests derived from CMB
observations.  Standard cosmological models predict that CMB
radiation is linearly polarized. However, some cosmological models
predict in addition a possible circular polarization of CMB
radiation \cite{cooray03}. In order to better constrain the
cosmological model with observations, it is crucial to determine the
degree of polarization of not only the CMB radiation but also the
other astrophysical sources whose signals are mixed with it. For an
excellent review on CMB polarization, see \cite{kamion97}. We have
treated the application to the detection of linearly polarized
sources in CMB maps elsewhere \cite{paco09}, but it is known that
extragalactic radio sources can also indeed show circular
polarization \cite{kirk06}.  Besides, circular polarization occurs
in many other astrophysical areas, from Solar Physics \cite{tris07}
to interstellar medium \cite{cox07}, just to put a few examples.
Therefore, in this paper we aim to address the general case of
linear plus circular polarization of compact sources.

In the general case we have three images $Q$, $U$ and $V$. Different
approaches can be used to deal with detection/estimation of
point-like sources embedded in a noisy background. On the one hand,
one can try to get the source polarization amplitude $A$ directly on
the $P$-map. In this approach, we will consider one filter, obtained
through the MLE applied to the modulus distribution (ModF). On the
other hand, we can operate with three matched filters, each one on
$Q$, $U$ and $V$ followed by a quadratic fusion and square root
(FF). We are trying to compare the performance of the two techniques
for estimating the position and polarization amplitude of a compact
source. Of course, in the case we have only the map of the modulus
of a vector and the components are unknown, the FF cannot be applied
but we can still use the ModF. Finally, we also apply a
matched filter (MF) directly on the $P$-map and compare this simple
method with the two new methods introduced in this paper.

In Section~\ref{sec:methodology} we will develop the methodology for
the case of $M$ images, because of the possible interesting
applications to the general $M$-dimensional case and in particular to
the 3-dimensional case (polarization). In Section~\ref{sec:simresults}
we will show the results when applying these techniques to numerical
simulations of images of $Q$, $U$ and $V$ that are relevant for the
detection of compact polarized sources in Astrophysics. Finally, in
Section~\ref{sec:conclusions} we give the main conclusions.

\section{Methodology} \label{sec:methodology}

\subsection{The case of $M$ images} \label{sec:n_images}

To develop our methodology, let us consider $M$ images with
intensities $d_{j}(\vec{x})$ at each point $\vec{x}$ of their common
domain, $j= 1,\ldots,M$. Hidden in these images there is an unknown
number of signals.  In this work we consider signals with compact
support (`compact sources' hereinafter), as for example galaxies in
CMB polarization images.  For simplicity, let us consider the case
of a single compact source embedded in the images. We will assume a
linear model
\begin{equation} \label{eq:basic_model}
  d_{j}(\vec{x}) = A_{j} \tau (\vec{x}) + n_{j}(\vec{x}).
\end{equation}
In this equation, the source is characterized by amplitudes $A_{j}$
in each image and by a spatial profile $\tau (\vec{x})$ that is the
same for the $M$ images.  This last condition is satisfied when the
instrument resolution is much higher than the source scale . The
source is immersed in noise $n_{j}(\vec{x})$ that is Gaussian and
independently distributed with zero mean and dispersion
$\sigma(\vec{x})$. Then the distribution of $d_{j}(\vec{x})$ is
Gaussian with mean $ A_{j} \tau (\vec{x})$ and dispersion
$\sigma(\vec{x})$. We will consider that the noise is non-stationary
but the dispersion is the same for the different images at the same
pixel. These conditions are typical in polarization images. By
construction, the total polarization map $P(\vec{x})\equiv (\sum_j
d_{j}^{2}(\vec{x}))^{1/2}$ includes a source characterized by a
total polarization amplitude $A\equiv (\sum_j A_{j}^{2})^{1/2}$, the
modulus of the vector $(A_1,A_2,....,A_M)$.

\subsubsection{ModF on the $P$-map} \label{sec:NPF_Nmaps}

If the noise is distributed normally and independently, the 1-pdf of
$d_j$ at any point of each image is
\begin{equation}
f(d_j(\vec{x})| A_j)=\frac{1}{\sqrt{2\pi}\sigma(\vec{x})}
\exp{\left[{\frac{-(d_j(\vec{x})-A_j
\tau(\vec{x}))^2}{2\sigma^2(\vec{x})}}\right]}
\end{equation}
\noindent where $\tau(\vec{x})$ is the known source profile at the
corresponding point. Since the noise is independent between the
different images, the $M$-pdf can be written as
\begin{eqnarray} \label{eq:npdf}
f(d_1(\vec{x}),...,d_M(\vec{x})| A_1,...,A_M)=
%\nonumber
%\\
\frac{1}{{(2\pi)}^{M/2}\sigma^M(\vec{x})} \exp{\left[-\sum_j
{\frac{(d_j(\vec{x})-A_j\tau(\vec{x}))^2}{2\sigma^2(\vec{x})}}\right]}.
\end{eqnarray}
\noindent Next, we will derive the distribution of $P(\vec{x})$, the
modulus of $(d_1(\vec{x}),d_2(\vec{x}),....,d_M(\vec{x}))$. By
changing to $M$-dimensional spherical coordinates,
$(P,\theta_1,......,\theta_{M-1})$, we write at any point $\vec{x}$,
(in order to simplify the notation we write $P$ instead of
$P(\vec{x})$ )
\begin{eqnarray}
 f(P,\theta_1,...,\theta_{M-1}|A)=
  \frac{P^{M-1}}{{(2\pi)}^{M/2}\sigma^M}
%\nonumber
%\\
\exp{\left[-\frac{P^2+A^2\tau^2}{2\sigma^2}+\frac{PA\tau
\cos{\theta_{M-1}}}{\sigma^2}\right]} \times \nonumber
\\(\sin{\theta_{M-1}})^{M-2}(\sin{\theta_{M-2}})^{M-3}\ldots \sin \theta_2
\end{eqnarray}
\noindent This is the joint pdf of the modulus P and the
corresponding angles, note that $A\equiv (\sum_j A_{j}^{2})^{1/2}$
and we have multiplied by the Jacobian of the coordinate change.
Finally, by integrating on the angles
$\theta_1,......,\theta_{M-1}$, we find the pdf of the modulus $P$:
\begin{eqnarray}   \label{eq:pdf_modulus}
f(P|A) = \frac{P^{M/2}}{\sigma^2
(A\tau)^{(M-2)/2}}
%\nonumber \\
  \exp{\left[-(A^2\tau^2 + P^2)/2\sigma^2\right]}
  I_{\frac{M-2}{2}}\left(A\tau\frac{P}{\sigma^2}\right),
\end{eqnarray}
\noindent where $I_{(M-2)/2}$ is the modified Bessel function of the
corresponding order. When $M$=2 we obtain the Rice distribution
\cite{rice}.  In the case $\sigma=1$ with a general $M$,
distribution (\ref{eq:pdf_modulus}) is the non-central chi
distribution with $M$ degrees of freedom and $\lambda=A\tau$
\cite{meyer67,anderson81}.  This was expected since we have derived
the pdf of the modulus of a vector whose components follow
independent Gaussian distributions with mean $A_j\tau$.  If there is
no source, the previous formula defaults to
\begin{equation} \label{eq:old6}
  f(P|0) = \frac{M P^{M-1}}{2^{M/2} \sigma^M \Gamma(1+M/2)}
  \exp{\left[-P^2/2\sigma^2\right]},
\end{equation}
\noindent where $\Gamma $ is the gamma function. In the case
$\sigma=1$, (\ref{eq:old6}) is the chi distribution. In particular,
when $M$=2 we obtain the Rayleigh distribution \cite{papoulis}.

Now we assume that the $M$ images are pixelized with the same pixel
size; since the noise is independent pixel to pixel, the different
values of $P$ at each pixel, $P_i$, $i= 1,\ldots,N$, with $N$ the
number of pixels, follow the two distributions
\begin{eqnarray}
  f(P_1,..,P_N|0)  =
%\\
\prod_i \frac{M P_i^{M-1}}{2^{M/2}
    \sigma_i^M \Gamma(M/2+1)}
  \exp{\left[-P_i^2/2\sigma_i^2\right]} & \ \ & (H_o), \ \ \label{eq:null}  \\
%\end{equation}
%  \nonumber \\
%
%\begin{equation}
f(P_1,...,P_N|A)   =
%\nonumber \\
\prod_i \frac{P_i^{M/2}
    \exp{\left[-\frac{A^2 \tau_i^2 +
          P_i^2}{2\sigma_i^2}\right]}}{{\sigma_i^2
      (A\tau_i)^{\frac{M-2}{2}}}} I_{\frac{M-2}{2}}\left(A\frac{P_i
    \tau_i}{\sigma_i^2}\right) & \ \ & (H_1), \ \  \label{eq:H1}
\end{eqnarray}
\noindent being $H_o$ and $H_1$ the null (absence of source) and the
alternative (presence of source) hypotheses, respectively, and
$\tau_i$ the profile at the $i^{\mathrm{th}}$ pixel. The
log-likelihood is defined by
\begin{eqnarray} \label{eq:log_likelihood}
  l(A|P_1,...,P_N) = \log  {f(H_1)} = \nonumber \\
  -A^2\sum_i\frac{\tau_i^2}{2\sigma_i^2}
%\nonumber \\
- N(M-2)/2\,
  \log A+ \sum_i\log \left[I_{(M-2)/2}
    \left(A\frac{P_i\tau_i}{\sigma_i^2}\right)\right].
\end{eqnarray}
\noindent Where we have only written the terms wich depend on $A$.
The MLE of the amplitude, $\hat{A}$, can be obtained by maximizing
the previous expression and is given by the solution of the equation
\begin{equation} \label{eq:estimator_A}
  \hat{A} \ \sum_i\frac{\tau_i^2}{\sigma_i^2} =
  \sum_iy_i\frac{I_{M/2}(\hat{A}y_i)}{I_{(M/2-1)}(\hat{A}y_i)},\ \ \ y_i\equiv
  \frac{P_i\tau_i}{\sigma_i^2}.
\end{equation}
\noindent This equation can be interpreted as a non-linear filter
operating on the data (ModF). This method allows us to obtain the
MLE of $A$ given a total polarization map $P$.

\subsubsection{Filtered Fusion (FF)} \label{sec:FF_Nmaps}

In this case we consider separately each one of the $M$ images. We
call $d_{ji}$ the intensity in the $j^{\mathrm{th}}$ image at the
$i^{\mathrm{th}}$ pixel and $\sigma_i$ the dispersion at each pixel.
The pdf of the intensities in the $j^{\mathrm{th}}$ image is
\begin{eqnarray} \label{eq:pdfimage}
f(d_{j1},...,d_{jN}| A_j)=
%\nonumber
%\\
\frac{1}{{(2\pi)}^{N/2}\sigma_1.....\sigma_N} \exp{\left[-\sum_i
{\frac{(d_{ji}-A_j\tau_i)^2}{2\sigma_i^2}}\right]}.
\end{eqnarray}
\noindent The MLE for $A_j$ yields
\begin{equation}\label{eq:MF} \label{eq:old12}
\hat A_j=\displaystyle\frac{\sum_i d_{ji}\tau_i /\sigma_i^2}{\sum_i
\tau_i^2 /\sigma_i^2}
\end{equation}
\noindent This is equivalent to the application of the linear
matched filter MF, $\Phi $, operating on each image $d_j$, as given
by \cite{paco08}:
\begin{equation}   \label{eq:MF}
  \Phi (\vec{x})\propto \frac{\tau (\vec{x})}{\sigma^2(\vec{x})}.
\end{equation}
Then, with the $M$ filtered images $\hat A_j$ we make the non-linear
fusion $\hat A \equiv \big(\sum_j \hat A_j^2\big)^{1/2}$, in order
to estimate the source polarization amplitude. We call this
operation FF. Note that this method is clearly different from the
ModF, now we calculate the MLE of each component $A_j$, given the
values $d_{ji}$ of that image and finally compute the modulus $A$ of
that estimated vector, whereas with the ModF we calculate the MLE of
A given the polarization map P. For using the FF, we need to know
all the images. On the other hand, for using the ModF we only need
to know the total polarization map $P$.

\subsubsection{Matched Fiter (MF)} \label{sec:MF_Nmaps}

This is a naive approach, assuming that we only know the
$P$-map. We estimate $ A$ according to the expression

\begin{equation}\label{eq:MF}
\hat A=\displaystyle\frac{\sum_i P_i \tau_i /\sigma_i^2}{\sum_i
\tau_i^2 /\sigma_i^2}
\end{equation}

\noindent with $P_i$ the polarization data at each pixel.
This simple technique can be useful for comparison with the more
elaborated methods presented in the previous subsections.

\subsection{The case of three images} \label{sec:3_images}

Now, we shall assume that we have the same compact source in three
images $Q$, $U$, $V$ (this is the standard notation for Stokes
parameters), characterized by amplitudes $A_Q$, $A_U$, $A_V$ and a
profile $\tau (\vec{x})$ immersed in noise $n_{Q,U,V} (\vec{x})$ that
is Gaussian and independently distributed with zero mean and
dispersion $\sigma(\vec{x})$. In general, we will consider that the
noise is non-stationary. We will assume a linear model for the three
images
\begin{equation}
  Q,U,V(\vec{x}) = A_{Q,U,V}\tau (\vec{x}) + n_{Q,U,V}(\vec{x}).
\end{equation}
The $P$-map, $P(\vec{x})\equiv
(Q^2(\vec{x})+U^2(\vec{x})+V^2(\vec{x}))^{1/2}$, is characterised by
a source with amplitude $A\equiv (A_Q^2+A_U^2+A_V^2)^{1/2}$.

\subsubsection{ModF on the $P$-map} \label{sec:modf3}

In the case of three images, formulas (\ref{eq:log_likelihood}) and
(\ref{eq:estimator_A}) can be written as
\begin{eqnarray}
  l(A|P_1,...,P_N) = \log {f(H_1)} =
  -A^2\sum_i\frac{\tau_i^2}{2\sigma_i^2}
%\nonumber \\
- N\log A+
  \sum_i\log \left[\sinh
    \left(A\frac{P_i\tau_i}{\sigma_i^2}\right)\right].
  \label{eq:log_likelihood3}
\end{eqnarray}
\begin{equation} \label{eq:estimator_A3}
\hat{A}\sum_i\frac{\tau_i^2}{\sigma_i^2} + \frac{N}{\hat{A}}=
\sum_iy_i\coth(\hat{A}y_i),\ \ \ y_i\equiv
\frac{P_i\tau_i}{\sigma_i^2}.
\end{equation}
\noindent This equation can be interpreted as a non-linear filter
operating on the data which is the (ModF) for this particular case.

\subsubsection{Filtered fusion (FF)}  \label{sec:ff3}

In this case, we use the same MF operating on each image $Q$, $U$,
$V$, as given by equation (\ref{eq:old12}). Then, with the three
filtered images $Q_{MF}$, $U_{MF }$, $V_{MF }$ we make the non-linear
fusion $\hat A \equiv (Q_{MF}^2+U_{MF}^2+V_{MF}^2)^{1/2}$ pixel by
pixel.

\subsubsection{Matched Filter (MF)} \label{sec:MF_3maps}

We just apply (\ref{eq:MF}) to the case of three images.

\section{Simulations and results} \label{sec:simresults}

As commented in the introduction, since the estimation of the
intensity of polarized sources is of great interest in Astrophysics,
we will use an example taken from CMB Astronomy in order to illustrate
the performance of the techniques introduced in the previous
section.

In order to compare and evaluate the performance of the two filters,
we have simulated images of $16\times 16$ pixels with a pixel angular
size\footnote{The angle of the sky subtended by a pixel of the
  detector, for a given telescope.} of $3$ arcmin. We simulate the
$Q$, $U$ and $V$ components of the polarization as follows: each
component consists of Gaussian uncorrelated noise, plus a polarized
point source filtered with a Gaussian-shaped beam whose full width
half maximum (FWHM) is $14$ arcmin. This is a typical example of a CMB
polarization experiment\footnote{This particular choice of the pixel
  and beam sizes corresponds to the specifications of the $70$ GHz
  channel of the ESA's \emph{Planck} satellite.}. So the source
polarization components can be written as
\begin{equation}
  s_{Q,U,V}  \equiv  A_{Q,U,V} \exp \left[ -\frac{|\vec{x}|^2}{2\gamma^2}
    \right],
\end{equation}
\noindent where $\gamma$ is the beam dispersion (angular size) and we
assume in this formula that the source is centered at the origin. We
consider stationary noise,with zero mean and r.m.s. deviation
$\sigma=1$ in some unit system\footnote{For this example, we use arbitrary intensity
  units, since the quantity of interest for our purposes is the signal
  to noise ratio of the sources.}. We take values of $A_Q$, $A_U$ and
$A_V$ ranging from $0$ to $2.5$ with a step of $0.5$. The number of
simulations is $100$ for each combination of triplets of values of
$A_Q$, $A_U$ and $A_V$.  After carrying out the corresponding
simulations for $Q$, $U$ and $V$, we add them quadratically and take
the square root to calculate $P=\sqrt{Q^2+U^2+V^2}$, the total
polarization.

We assume that we do not know the exact position of the source in the
map and then we place it at random in the image. We have considered
images of $16\times 16$ pixels in order to do fast calculations. In
order to avoid border effects, we simulate and filter $24\times 24$
pixel patches and, after the filtering step, we retain only the
$16\times 16$ pixel central square.

We use three different filters: the FF, as described in
section~\ref{sec:ff3}), which consists in the application of the
matched filter to the images in $Q$, $U$ and $V$ separately and then
the calculation/construction of the $P$-map from the
matched-filtered images, the ModF applied directly on $P$, derived
from the MLE applied to the modulus distribution and presented with
detail in section~\ref{sec:modf3} and a simple MF applied on
the $P$-map. We apply these filters to each simulation, centering
the filters successively at each pixel, since we do not know the
source position.  We estimate the source amplitude $A_{ModF}$ for
the ModF, in this case we calculate the value of $A$ which maximises
the log-likelihood, eq. (\ref{eq:log_likelihood3}). For the FF we
estimate separately and obtain
$A_{FF}=\sqrt{Q_{MF}^2+U_{MF}^2+V_{MF}^2}$. With the MF, we
obtain the estimator of A, (\ref{eq:MF}). Then, we have constructed
three maps ( one for the ModF , one for the FF and another one for
the MF) with the estimated values of $A$ at each pixel.

\begin{figure}[!t]
\centering
\includegraphics[width=\columnwidth]{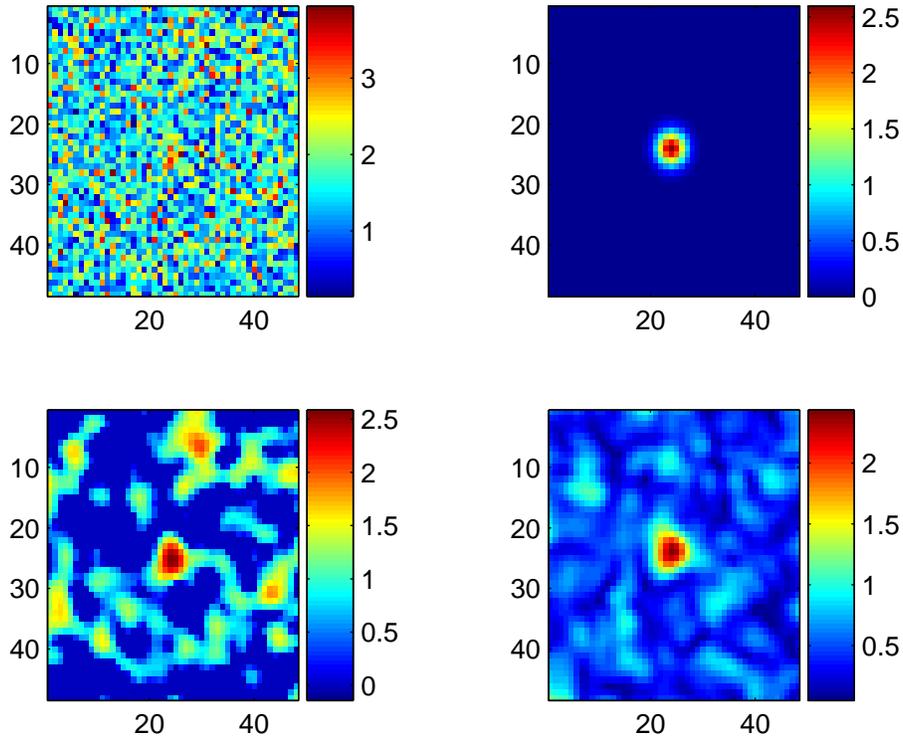}
\caption{Top left: Image of a polarized source filtered with a Gaussian
    beam. The source polarization components are
    $(A_Q,A_U,A_V)=(1.5,1.5,1.5)$ and it is embedded in stationary noise
    $\sigma=1$ . Top right: Image of the polarized source only. Bottom
    left: The first image after application of the ModF. Bottom right:
    The first image after the application of the FF.}
\label{fig:fig1}
\end{figure}

We show in Figure~\ref{fig:fig1} four images corresponding to a
polarized source with $(A_Q,A_U,A_V)=(1.5,1.5,1.5)$ embedded in
noise. For the sake of a better visualisation, we show $48\times48$
pixel images instead of the $16\times16$ sized images used in the
simulations. We show the original image in $P$ including noise and
source, the image of the source, the image filtered with the ModF
and finally, the image treated with the FF method. As we
will comment with more detail below, the performance of the MF is
much worse than that of the other filters and we will only show some
results obtained with the MF, leaving the figures and tables for the
comparison of the FF and the ModF. 

We compute the absolute maximum of each filtered map ($A_{ModF}$,
$A_{FF}$ and $A_MF$) and keep this value as the estimated
value of the polarization amplitude of the source and the position
of the maximum as the position of the source. Note that for the more
realistic case where more than one source can be present in the
images, it is still possible to proceed as described by looking for
local peaks in the filtered images.

We also calculate the significance level of each detection.  In
order to do this, we carry out 1000 simulations with $A_Q=0$,
$A_U=0$, $A_V=0$ and we calculate the estimated value of the source
polarization in this case for each filter. We consider the null
hypothesis $H_0$ (there is no polarized source) against the
alternative hypothesis $H_1$ (there is a polarized source). We set a
significance level $\alpha =0.05$; this means that we reject the
null hypothesis when a simulation has a estimated source
polarization amplitude higher than that of $95\%$ of the simulations
without polarized source. The previous significance fixes a lower
threshold ($A_*=1.99$ for the ModF and $A_*=1.17$ for the FF and
$A_*=3.81$ for the MF) defining a region of acceptance in
the space of polarization amplitudes. We define the power of the
test as $1-\delta$, with $\delta$ the probability of accepting the
null hypothesis when it is false, i.e. the power is the proportion
of simulations with polarized source with a estimated amplitude
higher than that of $A_*$. The higher the power the more efficient
the filter is for detection.

 We calculate the estimated value of the polarization amplitude
 $\hat{A}$ and compare it with the real value $A$, obtaining the
 relative error of the estimation and its absolute value for each
 simulation.We also calculate the estimated position of the detected
 source and obtain the position error expressed in terms of the number
 of pixels. The average of all these quantities for 100 simulations
 are presented in Table~\ref{tb:table1} for all the cases. The rows in
 the Table are sorted in ascending order of $A$.  We see in the table
 that the power for the FF is higher than for the ModF. The
 improvement is particularly high for $A$ values equal or lower than
 $2.5$. In this case the position and polarization errors are also
 lower for the FF. The FF can detect sources from $A\geq 1.8$ with
 power $\geq 0.99$ and average relative error (bias) $\leq 0.05$,
 average of its absolute value $\leq 0.14$ and average position error
 $\leq 0.53$. For $A =1.8$, the ModF detects with power $= 0.41$,
 average relative error $= 0.25$, average of its absolute value $=
 0.25$ and average position error $= 1.39$. For $A =1.8$, the MF
 detects with power $= 0.46$, average relative error $= 1.20$, average
 of its absolute value $= 1.20$ and average position error $=
 1.34$. The errors of the polarization estimation for the MF are much
 higher than for the other methods, whereas the power and the position
 error are similar to those obtained with the ModF.

The ModF and the FF perform in a similar way for signal-to noise ratio
$A \geq 3$, in this case the power is $1$ for both filters and the
bias is $\leq 0.02$. The ModF is also slower than the FF, due to the
maximization process involved in its application. However, note that
the FF cannot be applied when we only have the image of the modulus
i.e. we do not have information about the components, in this case the
ModF can be a suitable filter, specially for signal-to-noise ratio
$\geq 3$.

\begin{figure}[!t]
\centering
\includegraphics[width=\columnwidth]{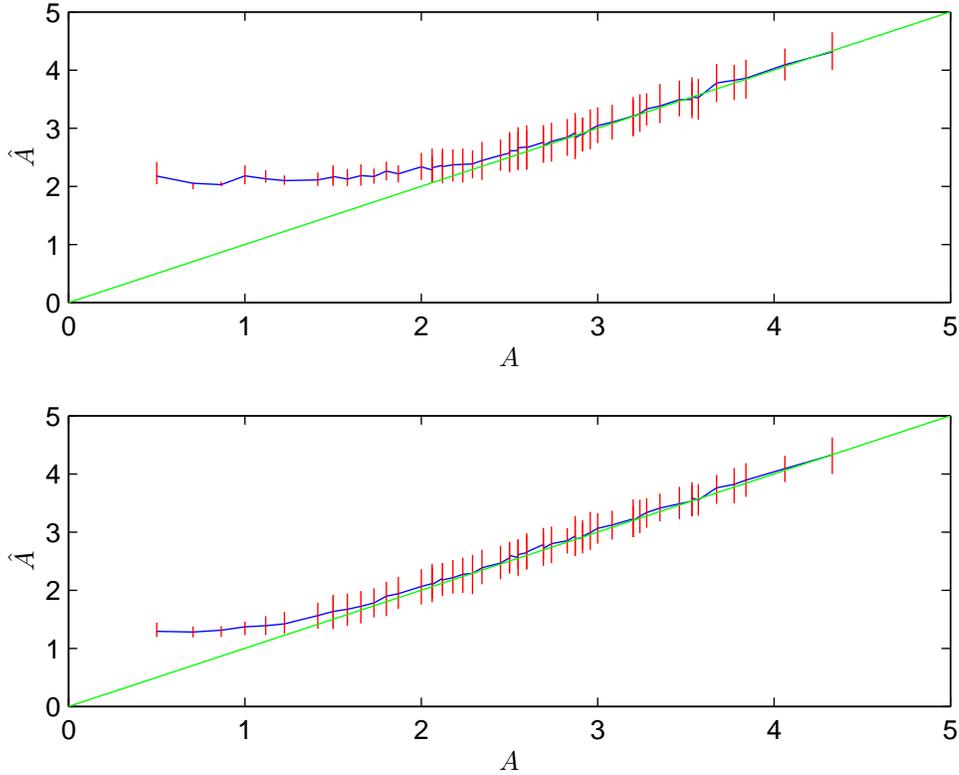}
\caption{Estimated source polarization amplitude $\hat{A}$ plotted
  against the real polarization amplitude $A$. The average and $68\%$
  confidence intervals of 100 simulations are plotted. Top : the ModF
  has been used. Bottom: the FF has been applied. The straight line
  $\hat{A}=A$ has been drawn for comparison.}
\label{fig:fig2}
\end{figure}

In Figure~\ref{fig:fig2}, we have plotted the estimated polarization
amplitude $\hat{A}$ against the real polarization amplitude $A$ for
the ModF and the FF. The average and $68\%$ confidence intervals of
100 simulations are shown. It is clear that the FF performs better
till $A\cong 3$. In Figure~\ref{fig:fig1}, we can also see the better
performance of the FF. From a qualitative point of view, the ModF
image shows more structure, with bright artifacts that could lead to
spurious detections, while the FF image looks smoother. This is easy
to understand if we think ModF is enhancing an inherently non-Gaussian
noise, whereas FF is the (non linear) composition of three smoothed
Gaussian noises.

\begin{figure}[!t]
\centering \includegraphics[width=\columnwidth]{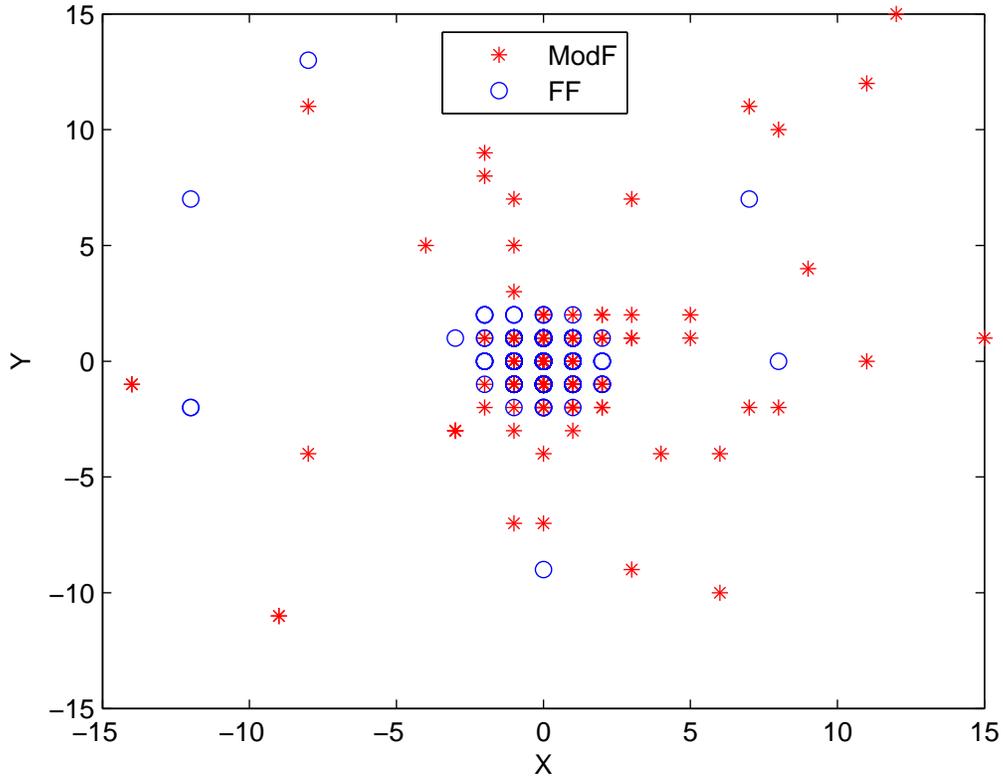}
\caption{Position error in number of pixels for the case $A_Q=0.5$,
  $A_U=1$, $A_V=1$. 1000 simulations have been considered.}
\label{fig:fig3}
\end{figure}

Finally, we have plotted in Figure~\ref{fig:fig3} the position error
in numbers of pixels for 1000 simulations in the particular case
$A_Q=0.5$, $A_U=1$, $A_V=1$. Only the simulations with a significance
$\alpha=0.05$ have been taken into account and represented. It is
clear that the position error is lower for the FF.

In order to try to understand the better performance of the FF, we
have carried out simulations of three-dimensional vectors whose
components are gaussian-distributed with dispersion $\sigma=1$ and
values of the mean ranging from $0.5$ to $2.5$. We have considered
the ModF, i.e we find the value of $A$ which maximizes
(\ref{eq:pdf_modulus}) with $\tau=1$ and the filtered fusion, which
amounts to averaging the components and calculating the modulus of
this average vector; this last method is also considered in
\cite{oberto} for the 2-dimensional case. The conclusion for these
simple cases is that the FF performs better than the ModF, specially
for the low signal-to-noise case, i.e. these simple examples confirm
our conclusions: it is more precise to estimate the modulus
estimating first the components and then calculating the square root
of the quadratic sum than calculating the MLE estimator directly on
the modulus.

\section{Conclusions} \label{sec:conclusions}

In this paper, we deal with the detection and estimation of the
modulus of a vector, a problem of great interest in general and in
particular in astrophysics when we consider the polarization of the
cosmic microwave background (CMB), extragalactic sources, the
interestellar medium or the Sun. The total polarization intensity $P$
is defined as $P\equiv (Q^2+U^2+V^2)^{1/2}$, where $Q$ , $U$ ( linear
polarization) and $V$ (circular polarization) are the Stokes
parameters. We consider the case of images in $Q$, $U$ and $V$
consisting of a compact source with a profile $\tau(\vec{x})$ immersed
in Gaussian uncorrelated noise. We intend to detect the source and
estimate its polarization amplitude by using three different methods,
two new and a standard matched filter. a) a filter operating on the
modulus $P$ (ModF) and based on the maximization of the corresponding
log-likelihood and b) a filtered fusion (FF) procedure, i.e. the
application of the matched filter (MF) on the images of $Q$, $U$ and
$V$ and the combination of the corresponding estimates by making the
non-linear fusion $\hat{A}\equiv (Q_{MF}^2+U_{MF}^2+V_{MF}^2)^{1/2}$
and c) a matched filter (MF) operating on the $P$-map. We
present the three filters in Section~\ref{sec:methodology} for the
general case of the modulus of a $M$-vector, and for the case of
three-dimensional vectors, deriving the corresponding expressions ,
eqs.  (\ref{eq:estimator_A}) and (\ref{eq:estimator_A3}), for the
estimation of the modulus of a vector with the ModF.

Since we are interested in the detection of polarized signals in
Astrophysics, we have only considered the three-dimensional vector
case in our simulations. Note, however that the methods can be
applied in the general case of combination of $M$ images to obtain the
image of the modulus of a vector.

We have compared the performance of the filters when applied to
simulated images consisting of the $Q$, $U$ and $V$ components of a
Gaussian-shaped signal with different intensities plus Gaussian
uncorrelated stationary noise. For each simulation and for the two
methods, we have estimated the source amplitude and position.
Besides, we have calculated the detection power for a fixed
significance $\alpha=0.05$ and the errors of the estimated amplitude
and position. We find that the performance of the FF is the best for
low signal-to-noise, the ModF performs like the FF for $A\geq 3$.
The MF produces very high errors in the polarization
estimation, making this filter unsuitable for the treated problem.

We want to point out the good performance of the ModF, which could
be an interesting alternative to the FF when we have an image of the
modulus, but we do not know the components of the vector.

Finally, it would be interesting to generalize our methods to the case
of images in which the noise at different pixels is not uniform or,
more generally, is correlated. This could be important for CMB
polarization observations. We leave this problem for further work.

%\begin{center}

\begin{table}
\caption{First column: triplets of values of $A_Q$, $A_U$ and $A_V$
  and the corresponding value of $A$ in $\sigma$ units.  Second and
  third columns: detection power for the ModF and FF (percentage) for
  a significance level $\alpha=0.05$. Fourth and fifth columns: flux
  relative errors for the two filters. Sixth and seventh columns:
  absolute value of the relative error. Eighth and ninth columns:
  position errors in numbers of pixels. In all the cases we have
  performed the average of 100 simulations.}
\label{tb:table1}
\centering
\scriptsize
\begin{tabular}{|c|c|c|c|c|c|c|c|c|}
 \hline
 & & & & & & & & \\
  \textbf{($A_Q,A_U,A_V;A$)}&  \textbf{pow$_{ModF}$}&
    \textbf{pow$_{FF}$} & \textbf{err$_{ModF}$}&  \textbf{err$_{FF}$} &
    \textbf{$|$err$|$$_{ModF}$} &  \textbf{$|$err$|$$_{FF}$} &
    \textbf{pos$_{ModF}$} &  \textbf{pos$_{FF}$} \\
& & & & & & & & \\
\hline
%\endfirsthead

% \hline
% & & & & & & & & \\
%  \textbf{($A_Q,A_U,A_V;A$)}&  \textbf{pow$_{ModF}$}&
%    \textbf{pow$_{FF}$} & \textbf{err$_{ModF}$}&  \textbf{err$_{FF}$} &
%    \textbf{$|$err$|$$_{ModF}$} &  \textbf{$|$err$|$$_{FF}$} &
%    \textbf{pos$_{ModF}$} &  \textbf{pos$_{FF}$} \\
%& & & & & & & & \\
%\hline
%\endhead
%
%\hline
%\endfoot

%\hline
%\endlastfoot

   \textbf{(0.00,0.00,0.50;0.50)} &  6 &  12 &  3.35 &  1.58 &  3.35 &
    1.58 &  4.36 &  3.83 \\ \textbf{(0.00,0.50,0.50;0.71)} &  1 &  22 &
    1.90 &  0.81 &  1.90 &  0.81 &  11.31 &  2.45
    \\ \textbf{(0.50,0.50,0.50;0.87)} &  7 &  41 &  1.34 &  0.52 &  1.34 &
    0.52 &  6.68 &  2.19 \\ \textbf{(0.00,0.00,1.00;1.00)} &  5 &  47 &
    1.18 &  0.37 &  1.18 &  0.37 &  3.80 &  1.99
    \\ \textbf{(0.00,0.50,1.00;1.12)} &  6 &  71 &  0.91 &  0.24 &  0.91 &
    0.24 &  3.44 &  1.23 \\ \textbf{(0.50,0.50,1.00;1.22)} &  6 &  76 &
    0.71 &  0.16 &  0.71 &  0.17 &  3.11 &  1.14
    \\ \textbf{(0.00,1.00,1.00;1.41)} &  17 &  89 &  0.49 &  0.10 &  0.49 &
    0.14 &  3.24 &  0.87 \\ \textbf{(0.00,0.00,1.50;1.50)} &  24 &  96 &
    0.45 &  0.09 &  0.45 &  0.15 &  2.34 &  0.70
    \\ \textbf{(0.50,1.00,1.00;1.50)} &  25 &  95 &  0.44 &  0.09 &  0.44 &
    0.15 &  2.17 &  0.65 \\ \textbf{(0.00,0.50,1.50;1.58)} &  27 &  96 &
    0.34 &  0.06 &  0.34 &  0.14 &  1.45 &  0.62
    \\ \textbf{(0.50,0.50,1.50;1.66)} &  27 &  97 &  0.32 &  0.04 &  0.32 &
    0.13 &  1.37 &  0.61 \\ \textbf{(1.00,1.00,1.00;1.73)} &  43 &  100 &
    0.25 &  0.03 &  0.25 &  0.11 &  1.53 &  0.70
    \\ \textbf{(0.00,1.00,1.50;1.80)} &  41 &  99 &  0.25 &  0.05 &  0.25 &
    0.14 &  1.39 &  0.43 \\ \textbf{(0.50,1.00,1.50;1.87)} &  50 &  100 &
    0.18 &  0.04 &  0.18 &  0.12 &  1.33 &  0.53
    \\ \textbf{(0.00,0.00,2.00;2.00)} &  59 &  100 &  0.17 &  0.03 &  0.17
    &  0.12 &  0.83 &  0.30 \\ \textbf{(0.00,0.50,2.00;2.06)} &  63 &  100
    &  0.12 &  0.02 &  0.13 &  0.12 &  0.72 &  0.40
    \\ \textbf{(1.00,1.00,1.50;2.06)} &  65 &  100 &  0.11 &  0.03 &  0.11
    &  0.10 &  0.92 &  0.41 \\ \textbf{(0.00,1.50,1.50;2.12)} &  76 &  100
    &  0.11 &  0.04 &  0.12 &  0.10 &  0.61 &  0.33
    \\ \textbf{(0.50,0.50,2.00;2.12)} &  74 &  100 &  0.10 &  0.03 &  0.12
    &  0.10 &  0.66 &  0.35 \\ \textbf{(0.50,1.50,1.50;2.18)} &  76 &  100
    &  0.09 &  0.02 &  0.11 &  0.11 &  0.73 &  0.33
    \\ \textbf{(0.00,1.00,2.00;2.24)} &  81 &  100 &  0.06 &  0.02 &  0.10
    &  0.10 &  0.62 &  0.33 \\ \textbf{(0.50,1.00,2.00;2.29)} &  81 &  100
    &  0.04 &  0.00 &  0.09 &  0.10 &  0.79 &  0.31
    \\ \textbf{(1.00,1.50,1.50;2.35)} &  86 &  100 &  0.04 &  0.02 &  0.11
    &  0.10 &  0.90 &  0.22 \\ \textbf{(1.00,1.00,2.00;2.45)} &  85 &  100
    &  0.03 &  0.01 &  0.09 &  0.09 &  0.56 &  0.24
    \\ \textbf{(0.00,0.00,2.50;2.50)} &  96 &  100 &  0.04 &  0.04 &  0.10
    &  0.09 &  0.51 &  0.18 \\ \textbf{(0.00,1.50,2.00;2.50)} &  96 &  100
    &  0.03 &  0.02 &  0.10 &  0.09 &  0.50 &  0.25
    \\ \textbf{(0.00,0.50,2.50;2.55)} &  95 &  100 &  0.04 &  0.02 &  0.11
    &  0.09 &  0.48 &  0.24 \\ \textbf{(0.50,1.50,2.00;2.55)} &  94 &  100
    &  0.03 &  0.00 &  0.10 &  0.10 &  0.68 &  0.20
    \\ \textbf{(0.50,0.50,2.50;2.60)} &  99 &  100 &  0.03 &  0.02 &  0.10
    &  0.09 &  0.58 &  0.17 \\ \textbf{(1.50,1.50,1.50;2.60)} &  95 &  100
    &  0.03 &  0.02 &  0.11 &  0.10 &  0.62 &  0.17
    \\ \textbf{(1.00,1.50,2.00;2.69)} &  99 &  100 &  0.02 &  0.03 &  0.10
    &  0.08 &  0.49 &  0.16 \\ \textbf{(0.00,1.00,2.50;2.69)} &  97 &  100
    &  0.00 &  0.01 &  0.09 &  0.08 &  0.50 &  0.14
    \\ \textbf{(0.50,1.00,2.50;2.74)} &  98 &  100 &  0.01 &  0.02 &  0.10
    &  0.09 &  0.38 &  0.15 \\ \textbf{(0.00,2.00,2.00;2.83)} &  100 &  100
    &  0.00 &  0.01 &  0.09 &  0.06 &  0.58 &  0.15
    \\ \textbf{(1.00,1.00,2.50;2.87)} &  99 &  100 &  -0.01 &  0.00 &  0.10
    &  0.08 &  0.44 &  0.17 \\ \textbf{(0.50,2.00,2.00;2.87)} &  98 &  100
    &  0.02 &  0.02 &  0.11 &  0.09 &  0.41 &  0.15
    \\ \textbf{(0.00,1.50,2.50;2.92)} &  100 &  100 &  -0.01 &  0.01 &
    0.08 &  0.06 &  0.32 &  0.09 \\ \textbf{(1.50,1.50,2.00;2.92)} &  100
    &  100 &  0.00 &  0.00 &  0.08 &  0.08 &  0.35 &  0.10
    \\ \textbf{(0.50,1.50,2.50;2.96)} &  100 &  100 &  0.01 &  0.01 &  0.09
    &  0.08 &  0.33 &  0.11 \\ \textbf{(1.00,2.00,2.00;3.00)} &  100 &  100
    &  0.01 &  0.02 &  0.08 &  0.07 &  0.27 &  0.07
    \\ \textbf{(1.00,1.50,2.50;3.08)} &  100 &  100 &  0.01 &  0.01 &  0.08
    &  0.06 &  0.27 &  0.05 \\ \textbf{(0.00,2.00,2.50;3.20)} &  100 &  100
    &  0.00 &  0.00 &  0.08 &  0.07 &  0.33 &  0.12 \\

\hline
\end{tabular}
\end{table}
%\end{center}

\section*{Acknowledgments}

The authors acknowledge partial financial support from the Spanish
Ministry of Education (MEC) under project ESP2004-07067-C03-01 and
from the joint CNR-CSIC research projects 2006-IT-0037 and 2008-IT-0059.

%\section*{References}

\bibliographystyle{elsarticle-num}
\bibliography{pol_arguesoetal_bib}

\end{document}